# The Importance of Song Context and Song Order in Automated Music Playlist Generation

Andreu Vall,[1] Massimo Quadrana,[2] Markus Schedl,[1] and Gerhard Widmer[1]

[1] *Department of Computational Perception, Johannes Kepler University, Austria*
[2] *Pandora Media, Inc., CA, U.S.A.*

andreu.vall@jku.at

## Abstract

The automated generation of music playlists can be naturally regarded as a sequential task, where a recommender system suggests a stream of songs that constitute a listening session. In order to predict the next song in a playlist, some of the playlist models proposed so far consider the current and previous songs in the playlist (i.e., the song context) and possibly the order of the songs in the playlist. We investigate the impact of the song context and the song order on next-song recommendations by conducting dedicated off-line experiments on two datasets of hand-curated music playlists. Firstly, we compare three playlist models, each able to consider a different song context length: a popularity-based model, a song-based Collaborative Filtering (CF) model and a Recurrent-Neural-Network-based model (RNN). We also consider a model that predicts next songs at random as a reference. Secondly, we challenge the RNN model (the only model from the first experiment able to consider the song order) by manipulating the order of songs within playlists. Our results indicate that the song context has a positive impact on the quality of next-song recommendations, even though this effect can be masked by the bias towards very popular songs. Furthermore, in our experiments the song order does not appear as a crucial variable to predict better next-song recommendations.

## Introduction

Automated music playlist generation is a specific task in music recommender systems in which the user receives a list of song suggestions that constitute a listening session, usually listened to in the given order. This is in contrast to the browsing scenario, in which users receive a collection of recommendations and actively choose their preferred option.

According to interviews with practitioners and postings to a dedicated playlist-sharing website, the choice of songs and their order–or at least their relative position–have been identified as important aspects when compiling a playlist (Cunningham et al., 2006). Although some approaches to playlist generation take the previous songs in the playlist (i.e., the song context) and the song order into consideration, to the best of our knowledge, they do not explicitly analyze the importance of these variables.

## Modeling Music Playlists

In this section we describe the models we use for automated music playlist generation. We adopt the following approach for every model. Two disjoint sets of playlists are available, one for training and one for test, such that all the songs in the test playlist also occur in the training playlists. Hyperparameter tuning, if necessary, is performed on a validation split that is withheld from the training set. Given one or several songs from a test playlist, a trained playlist model has to be able to rank all the candidate songs according to how likely they are to be the next song in the playlist.

**Song Popularity**

This model computes the frequency of each song in the training playlists. At test time, the candidate songs are ranked according to their frequency. Thus, the predictions of this model (equivalent to a unigram model–see e.g., Manning & Schütze (2000)) are independent of the current song.

**Song-Based Collaborative Filtering**

This is an item-based Collaborative Filtering (CF) model. A song $s$ is represented by the binary vector $\mathbf{p}_s$ indicating the playlists to which it belongs. The similarity of each pair of songs $s_i$, $s_j$ in the training set is computed as the cosine between $\mathbf{p}_{si}$ and $\mathbf{p}_{sj}$. At test time, the next-song candidates are ranked according to their similarity to the current song, but previous songs in the playlist are ignored.

**Recurrent Neural Networks**

Recurrent Neural Networks (RNNs) are a class of neural network models particularly suited to processing sequential data. They have a hidden state that accounts for the input at each time step while recurrently incorporating information from previous hidden states.

We adopt the approach proposed by Hidasi et al. (2016), where an RNN model with one layer of gated recurrent units is combined with a loss function designed to optimize the ranking of next-item recommendations. At test time, given the current and all the previous songs in the playlist, the RNN outputs a vector of song scores that is used to rank the next-song candidates.

## Playlist Datasets

The "AotM-2011" dataset (McFee & Lanckriet 2012) is a collection of playlists derived from the playlist-sharing platform Art of the Mix (www.artofthemix.org). Each playlist is represented by song titles and artist names, linked to the corresponding identifiers of the Million Song Dataset (MSD) (Bertin-Mahieux et al. 2011), where available.

The "8tracks" dataset is a private collection of playlists derived from 8tracks (https://8tracks.com), an on-line platform where users can share playlists and listen to playlists other users prepared. Each playlist is represented by song titles and artist names. Since there are many different spellings for the same song-artist pairs, we mimic the AotM-2011 dataset and use fuzzy string matching to resolve the song titles and artist names against the MSD.

A considerable number of playlists in the AotM-2011 contain songs by one or very few artists. In order to study more diverse playlists (which we assume to correspond to a more careful compilation process), we keep only the playlists with at least 3 unique artists and with a maximum of 2 songs per artist. Although the 8tracks dataset is not affected by this issue (the terms of use of the 8tracks platform require that no more than 2 songs from the same artist or album may be included in a playlist), we apply the same filters for the sake of consistency. Furthermore, we keep only the playlists with at least 5 songs. This ensures a minimum playlist length, that is required to study the effect of the song position on model performance. Finally, songs occurring in less than 10 playlists are removed to ensure that the models have sufficient observations for each song.

We randomly assign 80% of the playlists to the training set and the remaining 20% to the test set. Note that full playlists are assigned to either split. At test time, the model deals with playlists that were never seen before. As in any recommendation task blind to item content, the songs that only occur in test playlists need to be removed because they can not be modeled at training time.

The filtered AotM-2011 dataset includes 17,178 playlists with 7,032 songs by 2,208 artists. The filtered 8tracks dataset has 76,759 playlists with 15,649 songs by 4,290 artists.

## Evaluation of Playlist Models

A trained playlist model is evaluated by repeating the following procedure over all the test playlists. We show the model the first song in a playlist. It then ranks all the candidate songs according to their likelihood to be the second song in that playlist. We keep track of the rank assigned to the actual second song and of the fact that this was a prediction for a song in second position. We then show the model the first and the second actual songs. The model has to rank all the candidate songs for the third position, having now more context. In this way, we progress until the end of the playlist, always keeping track of the rank assigned to the actual next song and the position in the playlist for which the prediction is made.

A perfect model would always rank the actual next song in the first position. A random model would, on average, rank the actual next song approximately in the middle of the list of song candidates. An extremely poor model would rank the actual next song in the last position. Note that the actual rank values depend on the number of candidate songs available.

Previous research has often summarized the ranking results in terms of recall at $K$, where $K$ is the length of the list of top next recommendations (see e.g., Hariri et al. (2012), Bonnin & Jannach (2014), Hidasi et al. (2016)). However, the proposed evaluation setting may be too pessimistic in the music domain (Platt et al. 2002, McFee & Lanckriet 2011), where songs other than the actual one may serve as valid playlist continuations. As a consequence, long lists of next-song candidates are needed to observe the model behavior. In order to better observe the performance of each model, we opt for analyzing the full distribution of predicted ranks, summarized by the first quartile, the median and the third quartile rank values (Figures 1 and 2). This approach also facilitates the comparison of different models.

## Song Context

Figure 1 displays the ranks attained by the actual next songs in the test playlists, given the predictions of the considered playlist models. The distributions of attained ranks are split by the position in the playlist for which the next-song prediction is made. The popularity-based model and the song-based CF model, which have no context and a context of 1 song, respectively, do not improve their predictions as they progress through the playlists. The RNN model, which is aware of the full song context, improves its performance as it progresses through the playlist.

Regarding the absolute model performance, it is worth noting that the popularity-based model and the RNN model show comparable overall performances, despite the fact that the RNN model is much more complex. We could explain this apparent contradiction in our previous work (Vall et al. 2017) as an effect of the bias towards popular songs, common in the music domain. We found that the popularity-based model performs outstandingly well on the most popular songs, but performs poorly on the infrequent songs. On the other hand, the performance of the RNN model is unaffected by the song popularity.

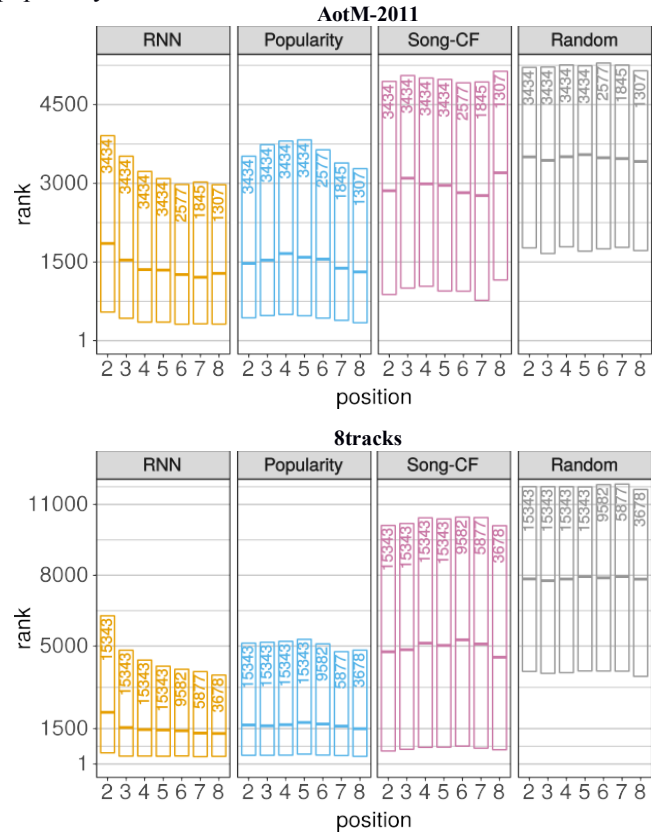

**Figure 1.** Distribution of ranks attained by the actual next songs in the test playlists (lower is better) for the AotM-2011 (top) and the 8tracks (bottom) datasets. Each panel corresponds to a playlist continuation model. The *x* axis indicates the position in the playlist for which a prediction is made. The *y* axis indicates the attained ranks and its scale relates to the number of songs in each dataset. The boxplots summarize the distribution of attained ranks by their first quartile, median and third quartile values. The number of rank values at every position is annotated.

## Song Order

We consider three song order manipulation experiments. For the first experiment we train the RNN model on original playlists, but we evaluate it on shuffled playlists (we refer to this setting as "shuffled test"). For the second experiment we train the RNN model on shuffled playlists and evaluate it on original playlists (we refer to this setting as "shuffled training"). Finally, we train and evaluate the RNN model on shuffled playlists (we refer to this setting as "shuffled training and test"). Figure 2 displays the ranks attained by the actual next songs in the test playlists, given the predictions of the RNN model under the different song order randomization experiments. The distributions of attained ranks are split by the position in the playlist for which the next-song prediction is made. The performance of the RNN model trained and evaluated on original playlists is kept as a reference.

The performance of the RNN model is comparable for all the song order randomization experiments, regardless of whether the song order is maintained, broken at test time or broken at training time. This result suggests that the song order may not be a crucial variable for automated music playlist generation. Even though we considered a competitive RNN model, further investigation on order-aware models is still required.

## Conclusion

In this work we explicitly analyzed the importance of considering the song context and the song order for automated music playlist generation. We conducted off-line experiments in two datasets of hand-curated music playlists, where we compared different playlist models with different capabilities. Our results indicate that the song context has a positive impact on next-song recommendations. Still, as we observed in previous works, the bias towards populars songs can mask the importance of considering the song context. On the other hand, the song order did not appear as a relevant variable to predict better next-song recommendations.

**Acknowledgements.** The authors want to thank Matthias Dorfer, Bruce Ferwerda, Rainer Kelz, Filip Korzeniowski, Rocío del Río and David Sears for helpful discussions. This research has received funding from the European Research Council (ERC) under the European Union's Horizon 2020 research and innovation programme under grant agreement No 670035 (Con Espressione).

## References

Bertin-Mahieux, T., Ellis, D. P., Whitman, B., & Lamere, P. (2011). The million song dataset. In Proceedings ISMIR.

Bonnin, G., & Jannach, D. (2014). Automated generation of music playlists: Survey and experiments. ACM Computing Surveys, 47(2), 1–35.

Cunningham, S. J., Bainbridge, D., & Falconer, A. (2006). "More of an art than a science": Supporting the creation of playlists and mixes. In Proceedings ISMIR.

Hariri, N., Mobasher, B., & Burke, R. (2012). Context-aware music recommendation based on latent topic sequential patterns. In Proceedings RecSys.

Hidasi, B., Karatzoglou, A., Baltrunas, L., & Tikk, D. (2016). Session-based recommendations with recurrent neural networks. In Proceedings ICLR.

Manning, C. D., & Schütze, H. (2000). Foundations of statistical natural language processing. MIT Press.

McFee, B., & Lanckriet, G. R. (2011). The natural language of playlists. In Proceedings ISMIR.

McFee, B., & Lanckriet, G. R. (2012). Hypergraph models of playlist dialects. In Proceedings ISMIR.

Platt, J. C., Burges, C. J., Swenson, S., Weare, C., & Zheng, A. (2002). Learning a Gaussian process prior for automatically generating music playlists. In Proceedings NIPS.

Vall, A., Schedl, M., Widmer, G., Quadrana, M., & Cremonesi, P. (2017). The importance of song context in music playlists. In RecSys 2017 Poster Proceedings.

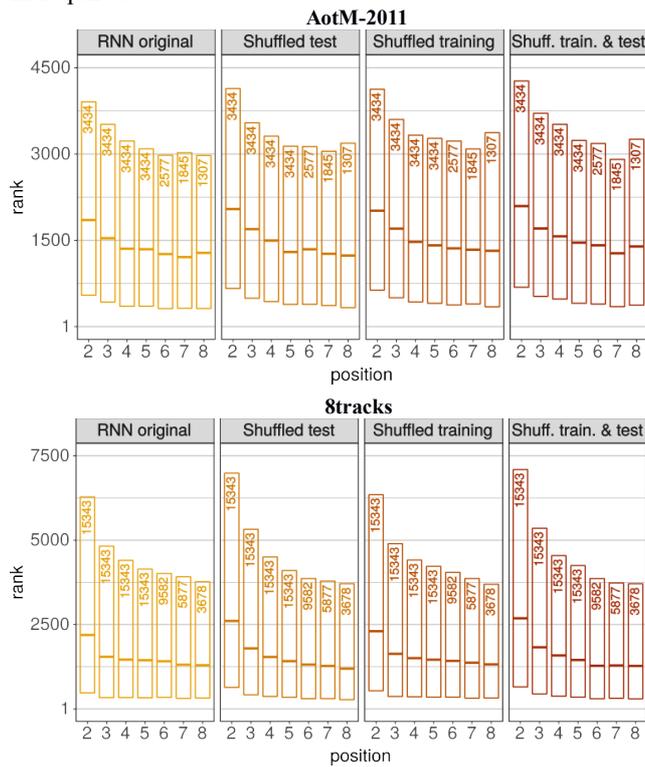

**Figure 2.** Distribution of ranks attained by the actual next songs in the test playlists (lower is better) for the AotM-2011 (top) and the 8tracks (bottom) datasets. The panels include the predictions of the RNN on the original playlists and on the song order randomization experiments. The *x* axis indicates the position in the playlist for which a prediction is made. The *y* axis indicates the attained ranks and its scale relates to the number of songs in each dataset. The boxplots summarize the distribution of attained ranks by their first quartile, median and third quartile values. The number of rank values at every position is annotated.